 \documentclass[aps,apl,twocolumn,superscriptaddress,amsmath]{revtex4-1} 
\usepackage{color} 
\usepackage{graphicx}   
\usepackage{xspace} 
\begin{document} 
 
\def\Gin{\Gamma_\mathrm{in}} 
\def\Gout{\Gamma_\mathrm{out}} 
\def\Gs{\Gamma_\mathrm{S}} 
\def\Gd{\Gamma_\mathrm{D}} 
\def\Gc{\Gamma_\mathrm{C}} 
\def\Grel{\Gamma_\mathrm{rel}} 
\def\Gabs{\Gamma_\mathrm{abs}} 
\def\Gem{\Gamma_\mathrm{em}} 
\def\Gdet{\Gamma_\mathrm{det}} 
\def\tdet{\tau_\mathrm{det}} 
\def\Vtdeg{V_\mathrm{2DEG}} 
\def\Iqpc{I_\mathrm{QPC}} 
\def\Vqpc{V_\mathrm{QPC}} 
\def\Gqpc{G_\mathrm{QPC}} 
\def\Vsd{V_\mathrm{SD}} 
\def\Vgl{V_\mathrm{G1}} 
\def\Vgr{V_\mathrm{G2}} 
\def\aGl{\alpha_\mathrm{G1}} 
\def\aGr{\alpha_\mathrm{G2}} 
\def\Idqd{I_\mathrm{DQD}} 
\def\tc{t_\mathrm{C}} 
\def\mul{\mu_\mathrm{1}} 
\def\mur{\mu_\mathrm{2}} 
\def\mus{\mu_\mathrm{S}} 
\def\mud{\mu_\mathrm{D}} 
\def\Ec{E_\mathrm{C}} 
\def\Ecm{E_\mathrm{Cm}} 
\def\Ecl{E_\mathrm{C1}} 
\def\Ecr{E_\mathrm{C2}} 
\def\Vqsd{V_\mathrm{QPC}} 
\def\Vdsd{V_\mathrm{DQD-SD}} 
\def\Rs{R_\mathrm{S}} 
\def\Rf{R_\mathrm{F}} 
\def\Cc{C_\mathrm{C}} 
\def\Cf{C_\mathrm{F}} 
\def\Vout{V_\mathrm{out}} 
\def\Vn{V_\mathrm{n}} 
\def\Vn{I_\mathrm{n}} 
\def\fBW{f_\mathrm{BW}}

 
 
\newcommand*{\um}{\ensuremath{\,\mu\mathrm{m}}\xspace} 
\newcommand*{\nm}{\ensuremath{\,\mathrm{nm}}\xspace} 
\newcommand*{\mm}{\ensuremath{\,\mathrm{mm}}\xspace} 
\newcommand*{\m}{\ensuremath{\,\mathrm{m}}\xspace} 
\newcommand*{\sqm}{\ensuremath{\,\mathrm{m}^2}\xspace} 
\newcommand*{\sqmm}{\ensuremath{\,\mathrm{mm}^2}\xspace} 
\newcommand*{\squm}{\ensuremath{\,\mu\mathrm{m}^2}\xspace} 
\newcommand*{\psqm}{\ensuremath{\,\mathrm{m}^{-2}}\xspace} 
\newcommand*{\psqmV}{\ensuremath{\,\mathrm{m}^{-2}\mathrm{V}^{-1}}\xspace} 
\newcommand*{\cm}{\ensuremath{\,\mathrm{cm}}\xspace} 
 
\newcommand*{\nF}{\ensuremath{\,\mathrm{nF}}\xspace} 
\newcommand*{\pF}{\ensuremath{\,\mathrm{pF}}\xspace} 
 
\newcommand*{\emob}{\ensuremath{\,\mathrm{m}^2/\mathrm{V}\mathrm{s}}\xspace} 
\newcommand*{\edos}{\ensuremath{\,\mu\mathrm{C}/\mathrm{cm}^2}\xspace} 
\newcommand*{\mbar}{\ensuremath{\,\mathrm{mbar}}\xspace} 
 
\newcommand*{\A}{\ensuremath{\,\mathrm{A}}\xspace} 
\newcommand*{\nA}{\ensuremath{\,\mathrm{nA}}\xspace} 
\newcommand*{\pA}{\ensuremath{\,\mathrm{pA}}\xspace} 
\newcommand*{\fA}{\ensuremath{\,\mathrm{fA}}\xspace} 
\newcommand*{\uA}{\ensuremath{\,\mu\mathrm{A}}\xspace} 
 
\newcommand*{\Ohm}{\ensuremath{\,\Omega}\xspace} 
\newcommand*{\kOhm}{\ensuremath{\,\mathrm{k}\Omega}\xspace} 
\newcommand*{\MOhm}{\ensuremath{\,\mathrm{M}\Omega}\xspace} 
\newcommand*{\GOhm}{\ensuremath{\,\mathrm{G}\Omega}\xspace} 
 
\newcommand*{\Hz}{\ensuremath{\,\mathrm{Hz}}\xspace} 
\newcommand*{\kHz}{\ensuremath{\,\mathrm{kHz}}\xspace} 
\newcommand*{\MHz}{\ensuremath{\,\mathrm{MHz}}\xspace} 
\newcommand*{\GHz}{\ensuremath{\,\mathrm{GHz}}\xspace} 
\newcommand*{\THz}{\ensuremath{\,\mathrm{THz}}\xspace} 
 
\newcommand*{\K}{\ensuremath{\,\mathrm{K}}\xspace} 
\newcommand*{\mK}{\ensuremath{\,\mathrm{mK}}\xspace} 
 
\newcommand*{\kV}{\ensuremath{\,\mathrm{kV}}\xspace} 
\newcommand*{\V}{\ensuremath{\,\mathrm{V}}\xspace} 
\newcommand*{\mV}{\ensuremath{\,\mathrm{mV}}\xspace} 
\newcommand*{\uV}{\ensuremath{\,\mu\mathrm{V}}\xspace} 
\newcommand*{\nV}{\ensuremath{\,\mathrm{nV}}\xspace} 
 
\newcommand*{\eV}{\ensuremath{\,\mathrm{eV}}\xspace} 
\newcommand*{\meV}{\ensuremath{\,\mathrm{meV}}\xspace} 
\newcommand*{\ueV}{\ensuremath{\,\mu\mathrm{eV}}\xspace} 
\newcommand*{\neV}{\ensuremath{\,\mathrm{neV}}\xspace} 
\newcommand*{\peV}{\ensuremath{\,\mathrm{peV}}\xspace} 
 
\newcommand*{\T}{\ensuremath{\,\mathrm{T}}\xspace} 
\newcommand*{\mT}{\ensuremath{\,\mathrm{mT}}\xspace} 
\newcommand*{\uT}{\ensuremath{\,\mu\mathrm{T}}\xspace} 
 
\newcommand*{\ms}{\ensuremath{\,\mathrm{ms}}\xspace} 
\newcommand*{\s}{\ensuremath{\,\mathrm{s}}\xspace} 
\newcommand*{\us}{\ensuremath{\,\mathrm{\mu s}}\xspace} 
\newcommand*{\ns}{\ensuremath{\,\mathrm{ns}}\xspace} 
\newcommand*{\rpm}{\ensuremath{\,\mathrm{rpm}}\xspace} 
\newcommand*{\minute}{\ensuremath{\,\mathrm{min}}\xspace} 
\newcommand*{\degree}{\ensuremath{\,^\circ\mathrm{C}}\xspace} 
 
\newcommand*{\EqRef}[1]{Eq.~(\ref{#1})} 
\newcommand*{\FigRef}[1]{Fig.~\ref{#1}} 
 
\newcommand{\Ket}[1]{\vert  #1 \rangle} 
\newcommand{\Bra}[1]{\langle  #1 \vert} 
\newcommand{\MatEl}[3]{\langle  #1 \vert #2 \vert #3\rangle} 
\newcommand{\Amp}[2]{\langle  #1 \vert  #2 \rangle} 
\newcommand{\mpar}[1]{\marginpar{\small \it #1}} 
\newcommand{\Avg}[1]{\langle  #1  \rangle} 
\newcommand{\lb}{\left[} 
\newcommand{\rb}{\right]} 
\newcommand{\lp}{\left(} 
\newcommand{\rp}{\right)} 
\newcommand{\E}{{\cal E}} 
 
\newcommand{\be}{\begin{equation}} 
\newcommand{\ee}{\end{equation}} 
\newcommand{\bea}{\begin{eqnarray}} 
\newcommand{\eea}{\end{eqnarray}} 
\newcommand{\HH}{{\cal H}} 
\newcommand{\LL}{{\cal L}} 
\newcommand{\KK}{{\cal K}} 
\newcommand{\VV}{{\cal V}} 
\newcommand{\GG}{{\sf G}} 
\newcommand{\tr}{{\rm tr\/}\,} 
\newcommand{\p}{\partial} 
\newcommand{\la}{\langle} 
\newcommand{\ra}{\rangle} 
 
\newcommand{\addLL}[1]{\textcolor{blue}{#1}} 
\newcommand{\addSG}[1]{\textcolor{magenta}{#1}} 
\newcommand{\addMR}[1]{\textcolor{red}{#1}} 
 
\renewcommand{\phi}{\varphi} 
\renewcommand{\epsilon}{\varepsilon} 
\renewcommand{\vec}[1]{{\bf #1}}

\title{Ultra-narrow ionization resonances in a quantum dot under broadband excitation} 
 \author{S.~Gustavsson} 
\affiliation {Solid State Physics Laboratory, ETH Zurich, CH-8093 Zurich, 
 Switzerland} 
\affiliation {Research Laboratory of Electronics, Massachusetts Institute of Technology, Cambridge, MA 02139, USA}
\author{M.~S. Rudner} 
\affiliation {Department of Physics, Harvard University, 17 Oxford St., Cambridge, MA 02138, USA} 
\author{L.~S. Levitov} 
\affiliation {Department of Physics, Massachusetts Institute of Technology, Cambridge, MA 02139, USA} 
 \author{R.~Leturcq} 
\affiliation {Solid State Physics Laboratory, ETH Zurich, CH-8093 Zurich, 
 Switzerland} 
\affiliation {Institute of Electronics, Microelectronics and Nanotechnologies, CNRS-UMR 8520, Department ISEN, 59652 Villeneuve d'Ascq, France} 
 \author{M.~Studer} 
\affiliation {Solid State Physics Laboratory, ETH Zurich, CH-8093 Zurich, 
 Switzerland} 
 \author{T.~Ihn} 
\affiliation {Solid State Physics Laboratory, ETH Zurich, CH-8093 Zurich, 
 Switzerland} 
 \author{K.~Ensslin} 
 \affiliation {Solid State Physics Laboratory, ETH Zurich, CH-8093 Zurich, 
 Switzerland} 
 
\begin{abstract} 

Semiconductor quantum dots driven by the broadband radiation fields of nearby quantum point contacts provide an exciting new setting for probing dynamics in driven quantum systems at the nanoscale. 
We report on real-time charge-sensing measurements of the dot occupation, which reveal sharp resonances in the ionization rate as a function of gate voltage and applied magnetic field.
Despite the broadband nature of excitation, the resonance widths are much smaller than the scale of thermal broadening. 
We show that such resonant enhancement of ionization is not accounted for by conventional approaches relying on elastic scattering processes, but can be explained via a mechanism based on a bottleneck process that is relieved near excited state level crossings. 
The experiment thus reveals a new regime of a strongly driven quantum dynamics in few-electron systems. 
The theoretical results are in good agreement with observations.

\end{abstract} 

\maketitle

Semiconductor quantum dots with proximal quantum point contacts (QPCs) are versatile systems in which a wealth of quantum dynamical phenomena can be realized and probed \cite{field:1993}.
In this work, we employ such a system to investigate ionization in a nanoscale artificial atom (a double quantum dot), using the QPC in a dual capacity as both a broadband emitter\cite{aguado:2000,onacQD:2006,gustavssonPRL:2007} and as a sensitive time-resolved charge detector \cite{vandersypen:2004, schleser:2004, fujisawa:2004,elzermanNature:2004,amasha:2008, Flindt:2009}.

Ionization is the process through which a bound electron in a quantum system is ejected to the continuum. 
Typically, ionization is a {\it threshold} process, turning on sharply when the quantum of energy in the excitation source
exceeds the electronic binding energy.
Additional structure in the above-threshold ionization rate may also appear at particular values of the excitation energy 
due to the presence of quasi-bound excited states (resonances). 
Such resonances are widely seen in atomic \cite{madden:1963, biondi:1979}, molecular \cite{madden:1982}, and nanoscale solid state systems \cite{blick:1995, oosterkamp:1997}.
However, when the excitation source has a broad power spectrum, all sharp features of the ionization spectrum are expected to be smeared out. 

\begin{figure}[htb]
\centering 
\includegraphics[width=0.9\linewidth]{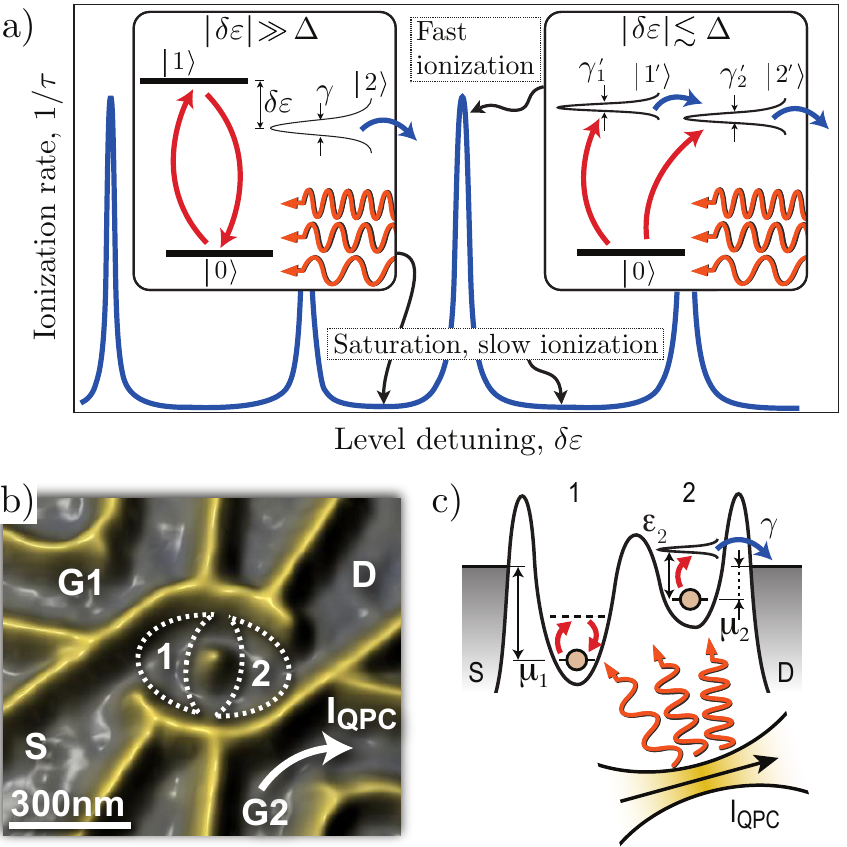} 
\caption{ Resonant enhancement of ionization at a level crossing. 
a) Broadband radiation from a QPC strongly couples the ground state $\Ket{0}$ to an excited state $\Ket{1}$. 
A bottleneck occurs because electron escape to the leads must take place through a different excited state, $\Ket{2}$. 
Near the level crossing, the states $\Ket{1}$ and $\Ket{2}$ hybridize to form new states $\Ket{1'}$ and $\Ket{2'}$, Eq.(\ref{eq:rotation}).
Both states couple to the leads, thus relieving the bottleneck.
b) AFM-image of the sample. The structure consists  
of two quantum dots (marked by 1 and 2) strongly coupled to a sensor/emitter QPC. Each QD contains a few tens of electrons.  
c) Schematic showing how the model in panel a) arises in a DQD. 
The many-body excited states $\Ket{1}$ and $\Ket{2}$ are distributed in both dots, with  
$\Ket{2}$ localized mostly in dot 2 offering the primary coupling to the leads.  } 
  \label{fig:Mechanism} 
\vspace{-5mm} 
\end{figure} 

%
In contrast to the picture above, in our experiment we find sharp resonances in the ionization rate 
as a function of gate voltages and external magnetic field.
We attribute these features to pairs of excited states that are swept through level crossings when the external fields are varied. 
Strikingly, even though the radiation is broadband, the observed linewidths are very narrow:
converting to an energy scale, we estimate the 
narrowest lines to be significantly narrower than the thermal broadening $k_{\rm B}T$ of electron energies in the leads \cite{Holleitner:2002, vanderwiel:2003}. 

We stress that the sharp resonances observed in our experiment are of a very different nature from those known e.g.~in resonant tunneling in double dots (c.f.~Ref.~\cite{vanderwiel:2003}). 
In our case, resonances appear in a photon-assisted inelastic transport regime, when pairs of {\it excitation energies} become degenerate; they do not require an absolute alignment of levels in the two dots, and remain sharp even for a broadband distribution of photon energies.
%
Furthermore, the observed resonant enhancement of ionization is not accounted for by models relying on perturbative scattering through the excited states. 
As discussed in greater detail below, such models predict, quite generally, ionization rates which are independent of level detuning.


To explain the phenomenon, we argue that the resonances 
arise from a new mechanism, which relies on a bottleneck process that is relieved near the level crossing [\FigRef{fig:Mechanism}a]. 
The essential ingredients of the model are the existence of a short-lived excited state with strong tunnel coupling to a reservoir, and another state, which is strongly coupled to the ground state by microwave excitation from the QPC.  
Coupling between these states near a level crossing  
eliminates a bottleneck for ionization, resulting in a sharp enhancement of the electron escape rate.  
Crucially, the resonances 
appear only when the interlevel transitions are strongly driven, near saturation. 
This is consistent with the observed power-dependence of the experimental traces (see below).

As illustrated in Figs.~\ref{fig:Mechanism}b and \ref{fig:Mechanism}c, electronic transitions are triggered by non-equilibrium fluctuations emitted from the voltage-biased QPC \cite{devoret:1990, girvin:1990}, leading to ionization of the DQD system which we detect in real time by monitoring the conductance of the same QPC \cite{gustavssonNWPRB:2008}.  
To bring the system into the regime where controlled ionization occurs and where the ionization rate can be measured, we reduce the tunnel couplings between the QDs and source and drain leads to a few kHz.  This ensures that the electron dwell times on and off the QDs are longer than the time resolution of the detector ($\tau_\mathrm{det} \sim 50 \us$), thus enabling real-time counting of tunnneling events.

\begin{figure}[tb] 
\centering 
\includegraphics[width=0.9\linewidth]{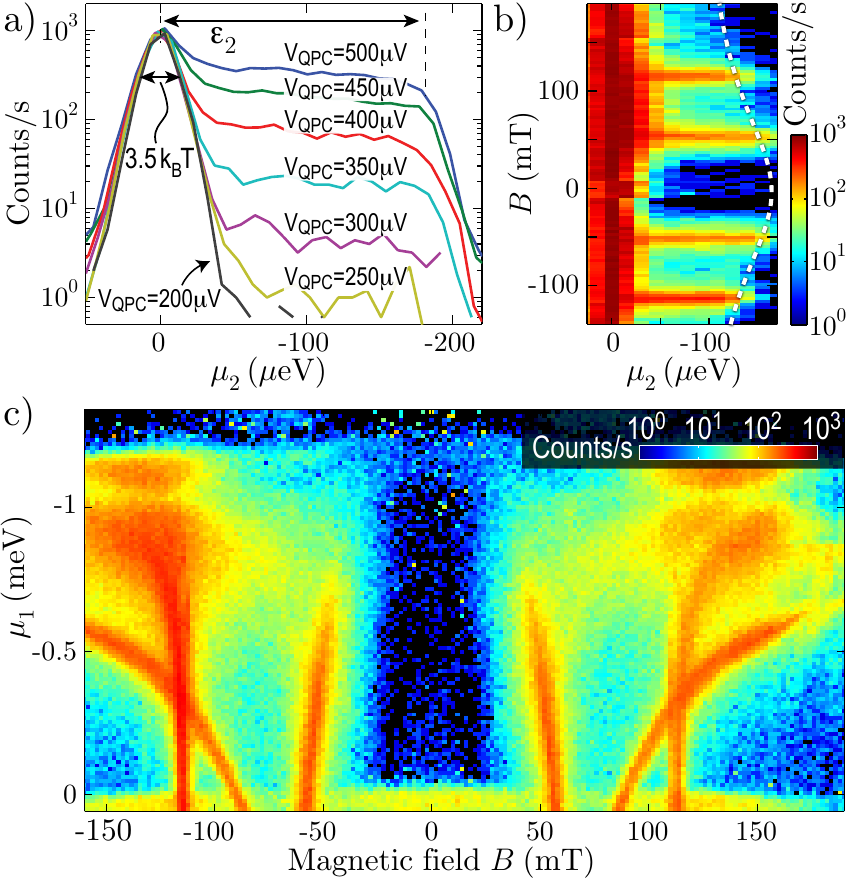} 
\caption{ Ionization measurements of a double quantum dot. 
 a) Number of electrons tunneling in and out of the dot per second, measured as a function of the electrostatic potential of dot 2 ($\mu_2$), relative to the Fermi energy of the drain, for different $\Vqpc$. 
The data is taken at $B=0\T$, with the potential of dot 1 is fixed at $\mu_1=-500\ueV$
(the values $\mu_{1,2}$ refer to the ground state levels and are obtained from the known capacitive lever arms of the gates \cite{vanderwiel:2003}).
 b) Count rate versus $\mu_2$ and magnetic field, measured at $\Vqsd = 350~\mathrm{\mu V}$ and $\mu_1=-400\ueV$, with  dot 1 containing one more electron than in panel a).  
The region $\mu_2<0$ exhibits sharp resonances as a function of magnetic field. 
c) Ionization rate as a function of magnetic 
field and dot potential $\mul$, with $\mu_2=-90~\mathrm{\mu eV}$. 
} 
 \label{fig:Rates} 
\end{figure}

In \FigRef{fig:Rates}a, we plot the count rate of electrons tunneling into and out of the dot 
as a function of the potential $\mu_2$ of dot 2 relative to that of the drain lead, measured for several values of $\Vqpc$.  
The peak at $\mu_2=0$ is due to equilibrium tunneling back and forth between dot 2 and the drain, with the peak height determined by the tunnel coupling and the peak width $3.5\, k_{\rm B} T$ set by the temperature $T = 90\mK$ in the lead\cite{kouwenhoven:1997}.  
For $|\mu_2| \gg k_{\rm B} T$, equilibrium fluctuations are suppressed. 
However, fluctuations in the QPC current may also drive inelastic transitions in the DQD when the energy $e\Vqpc$ supplied by the QPC voltage bias exceeds the required excitation energy \cite{onacQD:2006, gustavssonNWPRB:2008}, giving rise to the broad ionization shoulder seen in \FigRef{fig:Rates}a for large values of $\Vqpc$.

Note that the height of the shoulder is the only feature in \FigRef{fig:Rates}a that depends on $\Vqpc$. 
Neither the width of the shoulder, corresponding to  
the excitation energy $\epsilon_2 = 180\ueV$ (see \FigRef{fig:Mechanism}c), nor the shape of the equilibrium peak at $\mu_2=0$, are influenced by $\Vqpc$.  
Furthermore, the shoulder only appears when $e\Vqpc$ is larger than $\epsilon_2$, consistent with the emission spectrum of the QPC\cite{aguado:2000}.  
In the appendix we show that only the rate for tunneling out of the QD depends on $\Vqpc$, thus confirming that the increased count rate originates from ionization by radiation emitted by the QPC.
 
Using this method for measuring the ionization rate, we now study the rich phenomena that emerge when the excited states of the DQD are tuned by perpendicular (out of plane) magnetic field, $B$, and gate voltages. 
Figure~\ref{fig:Rates}b shows the electron count rate versus magnetic field and $\mu_2$. 
Similar to \FigRef{fig:Rates}a, the bright vertical feature indicating strong tunneling 
for $\mu_2 \approx 0$ arises from equilibrium fluctuations between dot 2 and the drain contact, while features at $\mu_2<0$ (to the right) indicate inelastic ionization processes.  
At $B=0$, the ionization rate is low, displaying only a weak shoulder of enhanced tunneling. 
At other values of $B$, however, sharp peaks appear indicating a resonant enhancement of ionization. 
 
It is important to point out that resonances occur when the {\it excitation energies} in the two dots are equal, $\epsilon_1 = \epsilon_2$, irrespective of the absolute alignment of the levels.  
Thus these features generally 
would not show up as tunneling resonances in elastic transport through the dots. 

The results shown in \FigRef{fig:Rates}b are surprising, as both the widths of the resonances  
(as low as a few mT) and their separations involve magnetic field scales that are much smaller than the fields associated with a flux quantum threading either the ring enclosed by the QDs (120 mT) or one of the QDs (several hundred mT) \cite{cronenwett:1997}.  
Two features in \FigRef{fig:Rates}b are particularly illuminating.  
First, the magnetic field strongly affects the ionization rate within the inelastic shoulder, while having only a weak effect on the shoulder extent (marked by a dashed white line in \FigRef{fig:Rates}b).  
This is consistent with the schematic in Fig.\ref{fig:Mechanism}c, provided that the energy level $\epsilon_2$ depends only weakly on $B$. 
Second, the equilibrium-tunneling peak at $\mu_2=0$ displays almost no $B$-field dependence. 
Thus, the resonant peaks in ionization cannot be explained by a $B$-field-induced modulation of the tunnel coupling between a single QD level and the lead. 

Further insight into the origin of the resonances can be obtained by tuning the gate voltages, which alters the confining potential of the QDs and changes their excitation spectra. As shown in \FigRef{fig:Rates}c, upon sweeping both $\mul$ and the magnetic field, different resonances behave essentially independently from each other: some resonances shift strongly with $\mul$, while others shift weakly. Interestingly, two of the resonances cross near $B = \pm 115\mT$, displaying no signatures of an avoided crossing (see Appendix).
Figure \ref{fig:Parabola} shows the results of a similar measurement, this time obtained with one electron removed from dot 1.
Individual resonances shift with $B$-field and $\mul$ in a manner qualitatively similar to that of the resonances in \FigRef{fig:Rates}c, but because of a larger number of resonances, the overall picture is more complex.
We note that the non-equidistant spacing of the resonances and their B-field dependence make them conceptually different from the phonon-absorption reported in Ref.~\cite{Granger:2012}.
The observed response of the resonances to $\mul$ and the reshuffling of resonances upon recharging dot 1 suggest that the resonant features arise from excited states in both dots. 
An example of an energy level configuration leading to such a pattern of resonances is discussed in Appendix \ref{app:Crossing}.

\begin{figure}[tb] \centering 
\includegraphics[width=0.9\linewidth]{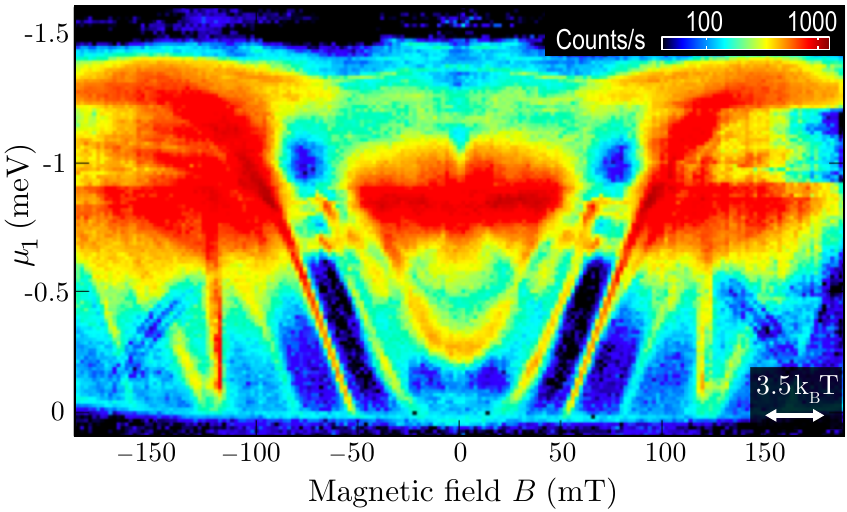} 
\caption{Ionization resonances as a function of magnetic field and dot potential $\mul$, 
measured for a different charge configuration than in \FigRef{fig:Rates}.  Many of the resonant features are significantly narrower than the thermal broadening of the electrons in the leads, as indicated by the scale bar in the lower-right corner of the figure. The conversion factor between energy and $B$-field is $0.8\meV/\mathrm{T}$.   
} 
\vspace {-0.4cm} 
\label{fig:Parabola} 
\end{figure} 

How narrow are the resonances? The narrowest peaks in \FigRef{fig:Parabola} have full-width half-maxima (FWHM) of about $3\mT$, which converted to energy gives an upper bound of $2.4\ueV$ 
(see Appendix \ref{app:Width}).  This is substantially lower than the width of the thermally-broadened peak in \FigRef{fig:Rates}a, which has a FWHM of $3.5 k_{\rm B} T = 27 \ueV$.   
To illustrate this comparison, we draw a scale bar in \FigRef{fig:Parabola} that corresponds to the FWHM of the thermally-broadened peak.

Below we show that ultra-narrow resonances 
can be understood within the simple model depicted schematically in Fig.\ref{fig:Mechanism}a. 
Before proceeding, 
it is important to point out that a simple perturbative calculation of the ionization rate does not account for the sharp resonances  when excited states are nearly degenerate.
Formally, this ``sum rule'' is illustrated as follows.
Consider three levels, $\Ket{0}$, $\Ket{1}$, and $\Ket{2}$, corresponding to the DQD ground state and two 
excited states. 
The state of the system $\Ket{\psi(t)}$ evolves according to $\left[i\frac{d}{dt} - H_0\right]\Ket{\psi} = V(t)\Ket{\psi}$, with 
\be
\label{H}
H_0 = \left(\begin{array}{cc} E_0 & {\bf 0} \\ {\bf 0} & {\bf H_{12}}\end{array}\right),\ V(t) = \alpha(t) \Big(\Ket{\phi_{12}}\Bra{0} + h.c.\Big). 
\ee
Here $\Ket{\phi_{12}} = C_1 \Ket{1} + C_2\Ket{2}$, with $|C_1|^2 + |C_2|^2 = 1$, and ${\bf H_{12}}$ is a $2\times 2$ (non-Hermitian) Hamiltonian accounting for the excited state energies, couplings, and decay rates to the leads (via imaginary level shifts).
Broadband radiation is described by 
$\overline{\alpha(t)\alpha(t')} = W_0\, \delta(t-t')$. 

Assuming the system is initialized in the state $\Ket{0}$ at time $t=0$ and setting $E_0 = 0$, we expand 
$\Ket{\psi(t)}$ as
$\Ket{\psi(t)} = \Ket{0} + \int_0^\infty dt' G_0(t - t')V(t')\Ket{0} + \cdots,$
with 
$G_0(t - t') = -ie^{-iH_0(t-t')}\,\theta(t-t')$. 
Keeping terms 
up to second order in $V(t)$, and averaging over all realizations of the broadband noise, the ionization rate $\Gamma(t) = -\frac{d}{dt}\log\overline{\Amp{\psi}{\psi}}$ is given by $\Gamma(t) = W_0\left[1 - \MatEl{\phi_{12}}{e^{i{\bf H}^\dagger_{\bf 12} t}e^{-i{\bf H_{12}}t}}{\phi_{12}}\right]$ (in the regime $W_0t \ll 1$ where the perturbative approach is valid).
For times longer than the intrinsic excited state lifetimes, the decay rate approaches a constant value $\bar{\Gamma} = W_0$, independent of the details of ${\bf H}_{12}$.

Consistency of the approach requires that the excited dot state populations must remain small, implying that 
the excitation must be weak compared with the smallest escape rate from the excited states.
In this case the ionization rate is controlled by coupling of the ground state to the excited states, which, under broadband excitation, is not sensitive to energy level detunings.
Thus the ionization resonances are not 
captured in this approach.

The bottleneck effect responsible for the ionization resonances 
appears when we consider the {\it population dynamics} of the 3 level system introduced above.
We illustrate the effect with a minimal model in which the broadband noise $V(t)$ primarily couples the ground state and one of the excited states, $\Ket{\phi_{12}} = \Ket 1$ in Eq.~(\ref{H}), while electron escape occurs from the other state, $\Ket{2}$. 
Dynamics within the excited state subspace are described by
\be
\label{eq:H} {\bf H_{12}} = \left(\begin{array}{cc}\epsilon_1 & \Delta/2\\ \Delta/2 & \epsilon_2 - i\gamma/2\end{array}\right),
\ee
where $\Delta$ describes the coupling between excited states $\Ket{1}$ and $\Ket{2}$, with energies $\epsilon_1$ and $\epsilon_2$, and $\gamma$ is the escape rate to continuum.
 %
This model describes the generic situation for our system, 
in which various states typically have very different characteristics.  
 %
For simplicity, we have set the direct excitation rate to state $\Ket{2}$ to zero.  
More generally, 
resonances appear as long as states $\Ket{1}$ and $\Ket{2}$ couple differently to the excitation source and to the leads. 

We investigate the behavior near level crossing $\delta\epsilon=\epsilon_1-\epsilon_2\approx 0$, 
taking into account the fact that the weak tunnel coupling regime realized in our system, with dwell times on a microsecond scale, is described by $\gamma\ll\Delta$. In this case, suppressing $\gamma$ and setting $V(t) = 0$, we diagonalize within the excited subspace spanned by $\Ket{1}$ and $\Ket{2}$ to obtain new hybridized eigenvectors 
\be\label{eq:rotation} 
\Ket{1'}=\alpha_1\Ket{1}+\beta_1\Ket{2} 
,\quad 
\Ket{2'}=\alpha_2\Ket{1}+\beta_2\Ket{2}. 
\ee 
This yields the eigenvalues $\epsilon'_1-i\gamma'_1/2$, $\epsilon'_2-i\gamma'_2/2$, where the decay rates are $\gamma'_1=|\beta_1|^2\gamma$,  $\gamma'_2=|\beta_2|^2\gamma$ (see Fig.~\ref{fig:Mechanism}a).  
The time-dependent field gives rise to nonzero transition rates from the ground state $\Ket{0}$ to the excited states $\Ket{1'}$ and $\Ket{2'}$, given by $w'_1=|\alpha_1|^2W$ and $w'_2=|\alpha_2|^2W$, where the net excitation rate $W$ is determined by the power spectrum of $V(t)$. 

When the detuning is large, $|\delta\epsilon|\gtrsim\Delta$, excitation occurs mainly to the non-decaying excited state; the $\Ket{0}-\Ket{1}$ transition may become saturated, with population transfer from the excited QD state to the continuum acting as a bottleneck for ionization. 
Near resonance, $|\delta\epsilon| \lesssim \Delta$, coupling between the excited states relieves the bottleneck and the ionization rate is enhanced. 
Note that when driving is weak, such that the excitation rate is small compared with $\gamma'_{1,2}$, excitation is the limiting step and no resonant enhancement is expected.

 \begin{figure}[t]
\centering 
\includegraphics[width=0.95\linewidth]{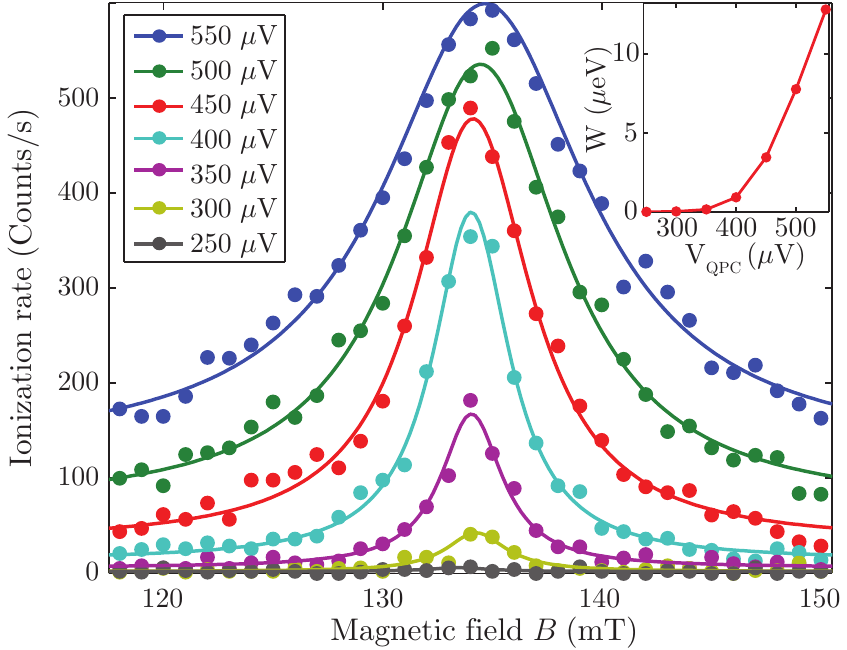} 
\caption{ Line shapes of the ionization resonances. 
Main panel: 
ionization rate as a function of magnetic field, measured for different values of $\Vqpc$.
Solid curves are fits to the model, Eqs.~(\ref{eq:H})-(\ref{eq:Gamma}), extended to include energy relaxation (see text). The resonance 
broadens when the excitation rate $W$ exceeds the coupling $\Delta$ between the excited states (at $\Vqpc>400\ueV$).
The fits include a small direct excitation rate ($0.5\%$ of $W$) between $\Ket{0}$ and $\Ket{2}$, to account for an increase in the background ionization level at high $\Vqpc$.
Inset: excitation rate 
$W$, extracted from the fits in the main panel. 
} 
  \label{fig:ModelRates} 
\end{figure} 

Using the excitation and decay rates defined above, we describe the dynamics of the populations $\vec P=(P_0,P_1,P_2)^{\rm T}$ 
of the three states via 
\be\label{eq:rate_eqs} 
\dot{\vec P}=-L\vec P 
,\quad 
L=\lp \begin{array}{ccc} 
w'_1+w'_2 & -w'_1 & -w'_2 \\ 
-w'_1 & w'_1+\gamma'_1 & 0 \\ 
-w'_2 & 0 & w'_2+\gamma'_2 
\end{array}\rp 
.
\ee 
The expected time  $\tau$ before ionization is given by $\tau=\int_0^\infty(P_0(t)+P_1(t)+P_2(t))dt$. 
Solving Eq.~(\ref{eq:rate_eqs}) as $\vec P(t)=e^{-Lt}\vec P(0)$, we have 
\be\label{eq:tau} 
\tau =(111)\lp \int_0^\infty  e^{-Lt}dt\rp (100)^{\rm T} =(111)L^{-1}(100)^{\rm T} 
. 
\ee 
From this expression we find that the net ionization rate $\Gamma(\delta\epsilon) = \tau^{-1}$ has a Lorentzian dependence on the detuning from level crossing (see \FigRef{fig:ModelRates}): 
\be\label{eq:Gamma} 
\Gamma(\delta\epsilon)=\frac{\Gamma_0}{\delta\epsilon^2+\gamma_*^2} 
,\quad 
\gamma_*^2=\Delta^2\frac{3W^2+\gamma^2+4W\gamma}{8W\gamma} 
, 
\ee 
with $\Gamma_0=\Delta^2 (W+\gamma)/8$. Strikingly, the width $\gamma_*$ of the Lorentzian (\ref{eq:Gamma}) 
is a nonmonotonic function of the excitation power $W$, 
diverging both in the limit of weak excitation, $W\ll\gamma$, and in the limit of strong excitation, $W\gg\gamma$.  
The narrowest resonance is realized when 
the excitation rate $W$ takes an optimal value such that the bottleneck for ionization is due to 
coupling between unhybridized states $\Ket{1}$ and $\Ket{2}$.  
Minimizing $\gamma_*$, we find $W=\gamma/\sqrt{3}$. 
In this case, the width of the resonance equals $\gamma_{\rm *\, min}=\Delta (3^{1/2}+2)^{1/2}/2\approx 0.97 \Delta$. 
 
In \FigRef{fig:ModelRates}, we plot the measured ionization rates for a resonance similar to the ones shown in \FigRef{fig:Rates} and \FigRef{fig:Parabola}, for several values of the QPC bias voltage (excitation power). 
The solid lines indicate fits to a modified form of \EqRef{eq:Gamma} which includes the effect of internal relaxation processes from the excited states back to the ground state (see Appendix).  
Such relaxation, which is not fundamental to the mechanism but appears needed for good quantitative agreement with the experimental observations, limits the efficiency of the ionization process while preserving the Lorentzian form of the resonances. 
For the fits, we assume the relaxation time $T_1 = 9\ns$ due to phonons \cite{fujisawa:2002,petta:2004} 
to be the same for both excited states, whereas $\gamma = 6\peV$ is known from the time-resolved measurement of the tunneling rate between the excited state in dot 2 and the drain lead. 
The fitting yields the same coupling $\Delta = 1.3 \pm0.1 \ueV$ independent of QPC bias voltage, as expected from the model. 
The coupling energy is consistent with values typical for resonant tunneling in quantum dot systems \cite{kouwenhoven:1997}.
 

In summary, we have discovered 
sharp resonances in the ionization rate of 
a quantum dot driven by broadband radiation. 
Ionization resonances arise due to a bottleneck process involving pairs of excited states that couple differently to a reservoir and to the microwave excitation, with the state more strongly coupled to the reservoir acting as a probe for other states. 
General arguments show that such resonances are only expected in a strong driving regime, where the perturbative description based on resonant tunneling between excited states breaks down.
The experiment utilizes the versatility of the coupled QD/QPC system, 
providing new means for probing strongly driven nanoscale systems.



We thank D. C. Driscoll and A. C. Gossard at Materials Department, University of California, Santa Barbara, California, USA, for fabricating the wafers used in this experiment. 

\appendix
\section{Methods} 
The device, pictured in Fig. 1b, was fabricated by local oxidation \cite{Fuhrer:2004} of a GaAs/Al$_{0.3}$Ga$_{0.7}$As heterostructure, containing a two-dimensional electron gas located 34 nm below the surface (mobility $3.5 \times 10^5~\mathrm{cm^2/Vs}$, 
density $4.6\times 10^{11}~\mathrm{cm}^{-2}$). 
The dots are coupled via two separate 
tunneling barriers, formed in the upper and lower arms between the 
dots.  The charging energy and the energy level spacing are about $1.3 \meV$ and $100-200 \ueV$ for each dot. From the geometry we estimate each QD to contain around 30 electrons.  We measured Aharonov-Bohm oscillations in transport to ensure that both barriers are open and have roughly equal tunnel coupling strength \cite{gustavssonNL:2008}. All measurements were performed in a dilution refrigerator with an electron temperature of 90 mK. 
 
In this work, we are tuning the excited state levels $\epsilon_1$ and $\epsilon_2$ by applying a perpendicular magnetic field.  Since $\epsilon_1$ and $\epsilon_2$ are defined relative to the ground state energies $\mul$ and $\mur$ (see Fig. 1), we first separately determined how $\mul$ and $\mur$ shift with $B$-field by measuring the resonant tunneling occurring when the ground states align with the Fermi levels in the leads.  For all B-field measurements presented in the paper, compensation voltages were applied to the gates \emph{G1}and \emph{G2}, to always keep the ground states aligned with the leads at $\mul,~\mur=0$.

\section{Tunneling rates for entering and leaving the QDs} 
\begin{figure}[hbt] 
\centering 
\includegraphics[width=\linewidth]{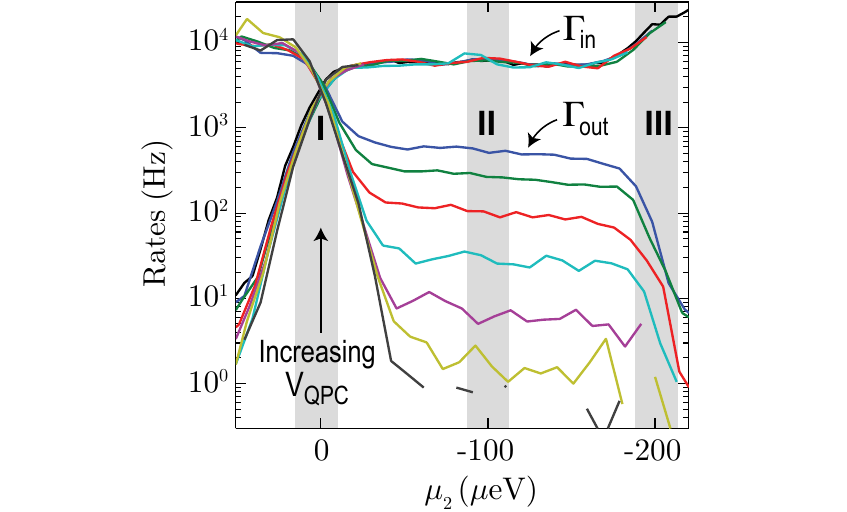} 
\caption{Rates for electrons tunneling into and out of the QD. The QPC bias is ranging from $\Vqsd = 200,~250,~300,~\ldots,~500\uV$.} 
\label{fig:SupRates} 
\end{figure}

In \FigRef{fig:SupRates}, we plot 
the rates $\Gin$ and $\Gout$ for electrons tunneling into and out of the QD. 
The rates were extracted from the same set of data as in Fig 2a in the main paper, taking the finite bandwidth of the detector into account \cite{naaman:2006}. 
At the position marked by I in \FigRef{fig:SupRates}, the tunneling is due to equilibrium fluctuations and the rates 
for tunneling into and out of the QDs are equal. In the regime of 
QD excitations (case II in \FigRef{fig:SupRates}), the rate related to absorption ($\Gout$) 
increases strongly with bias voltage over the QPC. 
Continuing to case III, when $|\mu_2|>\epsilon_2$ the excited state drops below the Fermi level of the 
source lead and the absorption rate drops quickly. At the 
same time, $\Gin$ increases as the refilling of an electron into QD2 
may occur through either the ground state or the excited state. 
The rate for tunneling into the $\Gin$ does not show any major  QPC bias 
dependence over the full range of the measurement. This is expected, since the refilling of an electron into the QDs does not require absorption of energy. 
  
\section{Magnification around crossing of resonances} 

\begin{figure}[hbt] 
\centering 
\includegraphics[width=\linewidth]{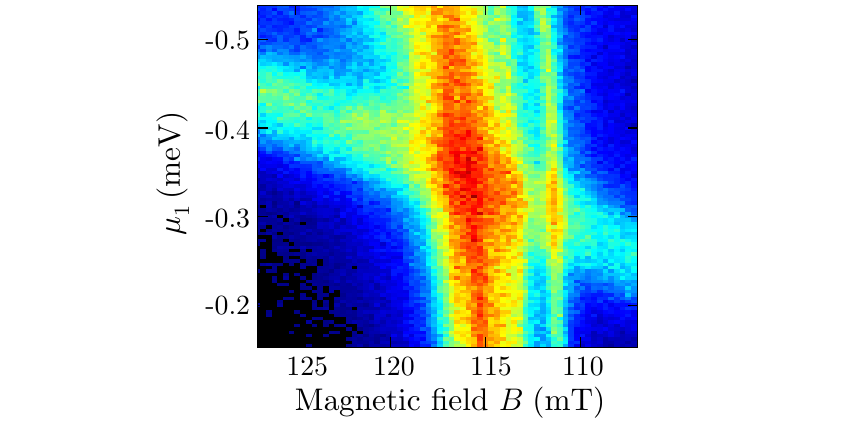} 
\caption{Magnification of a region in Fig.~2c in the main paper. The vertical line is split into two, with the finer structure 
 having a width below $1\mT$. The data is plotted on a linear linear color scale ranging from 0 to 500 counts/s. 
} \label{fig:SupZoom} 
\end{figure} 

Figure \ref{fig:SupZoom} shows a magnification of the region 
around the crossing of resonances in Fig.~2c in the main paper. The vertical feature is found to be split into two peaks, with the smaller sub-peaks having a full-width half maximum below $1\mT$. 
There is no anticrossing visible in the regime where the two main resonances meet. Within the resolution of the measurement, the two resonances appears to be uncoupled.

\section{Estimating the width of the resonances}\label{app:Width} 
The narrowest resonances seen in Fig.~3a in the main paper have a full-width half maximum (FWHM) of around $3\mT$.
To convert this width to an energy scale, we estimate the energy shift required to bring different states into resonance by changing the $B$ field. 
The orbital shift of the QD levels with $B$
is given by $\Delta E/B_0 \lesssim 0.4 \meV/\T$, where $\Delta E = 100-200\ueV$ is the level spacing and $B_0 = 500\mT$ is the magnetic field associated with a flux quantum threading one of the dots.  This is an upper bound for the shift, since hybridization of the orbital states generally leads to flattening of the bands \cite{fuhrer:2001}, but the value is consistent with the shift marked by the dashed line in Fig. 2b. 
For two states shifting in opposite directions, we estimate an upper bound of $0.8 \meV/\T$ for the conversion factor from magnetic field to energy. 
This yields an upper bound of $2.4\ueV$ for the FWHM of the narrowest features in Fig.~3, thus substantially lower than the width of the thermally-broadened peak in Fig. 2a, which has a FWHM of $3.5 k_{\rm B} T = 27 \ueV$.   
To illustrate this comparison, we draw a scale bar in Fig. 3 that corresponds to the FWHM of the thermally-broadened peak. 
Since the states shift differently with magnetic field, the scale bar only serves as a lower bound for the energy resolution due to the thermal broadening.  Still, it is clear that several of the resonances in Fig. 3 are considerably narrower than that lower bound.

\section{Model of energy levels giving the positions of the level crossings as a function of magnetic field and gate voltage} \label{app:Crossing}

In the main text we argue that the resonances correspond to level crossings between excited states in the two dots.
Here we describe a plausible configuration of energy levels which yields a similar pattern of resonances to that observed in the data.  Unfortunately, the data at hand does not provide enough information to uniquely determine the energy spectrum of the two quantum dots.  Instead, the purpose of this section is to show that a simple model involving a few excited states with uncomplicated gate voltage dependence is enough to re-create the fairly complex resonance curves seen in the experimental data.

We start with the experimental configuration and resonance data shown in Fig. 2c.  A simple scheme involves level crossings between a single excited state coupled to the lead (denoted $\Ket{2}$) and a set of three isolated excited states $\Ket{1a}$, $\Ket{1b}$, $\Ket{1c}$, with energies varying differently with gate voltage and magnetic field.
The resonances occur whenever the energy $\epsilon_2$ of state $\Ket{2}$ matches the energy of one of the other excited states.
 \begin{figure}[b] 
\centering 
\includegraphics[width=\linewidth]{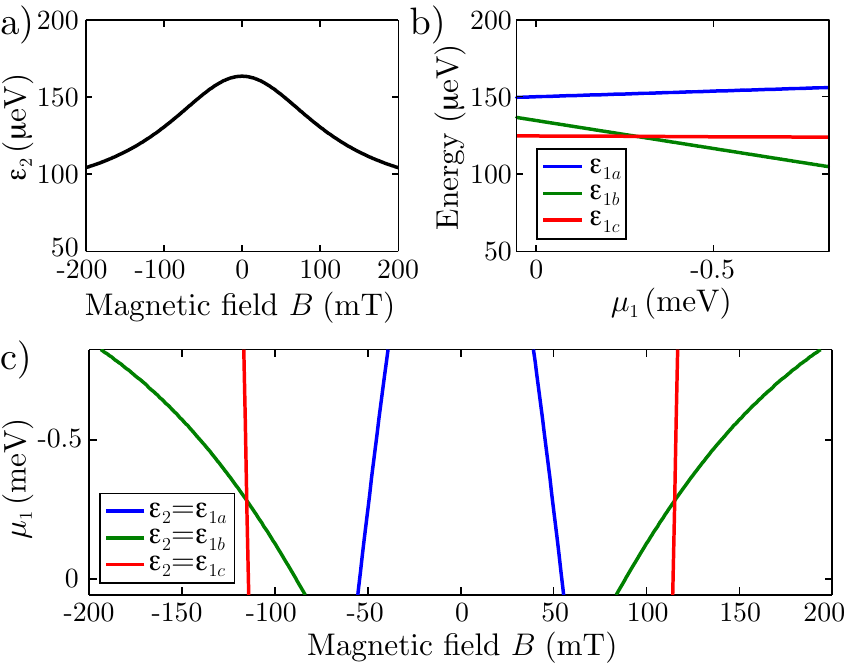} 
\caption{a) Magnetic field dependence of the energy $\epsilon_2$ of the state $\Ket{2}$ that is strongly coupled to the reservoir.  The curve is the same as the dashed line in Fig.~2b. 
b) Gate voltage dependence of the energies of states $\Ket{1a}$, $\Ket{1b}$ and $\Ket{1c}$, chosen so that panel c) reproduces the measured resonances in Fig.~2c. 
c) Position of the crossings between the state $\Ket{2}$ and three states $\Ket{1a}$, $\Ket{1b}$, $\Ket{1c}$ as a function of magnetic field and gate voltage, calculated using the magnetic field and gate voltage dependences in panels a) and b). 
} \label{fig:SupCrossings} 
\end{figure} 

 
 
In general, all excited states shift differently as a function of both the magnetic field and the gate voltages.   
We assume that the state $\Ket{2}$ is localized predominantly in dot 2, and that it is strongly coupled to the drain reservoir.
The energy $\epsilon_2$ of this state depends only very weakly on the potential $\mu_1$ that controls dot 1. This is consistent with the characteristics of our device (see Fig.~1 of the main text).  With these restrictions, the conditions for the resonances become 
\be\label{eq:resonanceCondition} 
\epsilon_2(B) = \epsilon_{1\alpha}(\mul,B),  
\ee 
where $\alpha = \{a, b, c\}$.  In the following, we are going to assume that the energies $\epsilon_{1\alpha}(\mul,B)$ are independent of $B$-field.  This assumption is not physically motivated, but rather servers to show that we can re-create the resonance data seen in the experiment with the simplest possible model.  With this simplification, Eq.~(\ref{eq:resonanceCondition}) becomes 
\be\label{eq:simple} 
\epsilon_2(B) = \epsilon_{1\alpha}(\mul).  
\ee 

The shape of $\epsilon_2(B)$  is known experimentally from the measurement in Fig. 2b in the main text (dashed line), which is reproduced in \FigRef{fig:SupCrossings}a.  
%
By combining the measured dependence of $\epsilon_2(B)$ with the conditions in \EqRef{eq:simple}, we can determine how the energies $\epsilon_{1\alpha}$ must shift with potential $\mul$ in order to produce the resonances seen in Fig.~2c in the main text.  The extracted values of $\epsilon_{1\alpha}(\mul)$ are plotted in \FigRef{fig:SupCrossings}b, and the resulting positions of the crossings $\epsilon_2 =\epsilon_{1a}$, $\epsilon_{1b}$, $\epsilon_{1c}$ as a function of magnetic field and gate voltage are shown in \FigRef{fig:SupCrossings}c.  

Despite the simplicity of the model, the curves reproduce the pattern of resonances in Fig.~2c of the main text. 
However, we stress again that the method does not provide any information about the $B$-field dependence of $\epsilon_{1a}$, ..., $\epsilon_{1c}$, and therefore only serves to show that a simple energy dependence is enough to re-create the complex resonance maps seen in the experimental data.  A similar approach can be used to re-create the resonance conditions for the data shown in Fig.~3 of the main text.

\begin{widetext} 
\section{Model of ionization resonances} 
In this section we present the rate equation model used to describe the resonant enhancement of ionization observed near excited level crossings. 
 In terms of the excitation and escape rates $w'_{1,2}$ and $\gamma'_{1,2}$ defined in the main text (below we suppress the primes for notational simplicity), and additional internal relaxation rates $\Gamma_{1,2}$ that describe relaxation from the excited states $\Ket{1}$ and $\Ket{2}$ to the ground state, the dynamics of the system is described by rate equations for the populations of the three levels: 
\be\label{eq:rate_eqs} 
\dot{\vec P}=-L\vec P 
,\quad 
L=\lp \begin{array}{ccc} 
w_1+w_2 & -(w_1 + \Gamma_1) & -(w_2 + \Gamma_2) \\ 
-w_1 & w_1+\gamma_1 + \Gamma_1 & 0 \\ 
-w_2 & 0 & w_2+\gamma_2 + \Gamma_2 
\end{array}\rp 
, 
\ee 
with $\vec P=(P_0,P_1,P_2)^{\rm T}$.  
 
The lifetime of the system, i.e.~the expected time before ionization, can be found from $\tau=\int_0^\infty(P_0(t)+P_1(t)+P_2(t))dt$. Solving the rate equations (\ref{eq:rate_eqs}) in terms of a matrix exponential as $\vec P(t)=e^{-Lt}\vec P(0)$, we have 
\be\label{eq:tau} 
\tau =(111)\lp \int_0^\infty  e^{-Lt}dt\rp (100)^{\rm T} =(111)L^{-1}(100)^{\rm T} 
. 
\ee 
Inverting the matrix $L$ and substituting the result into Eq.(\ref{eq:tau}), we obtain 
\be 
\tau =\frac{(w_1+\gamma_1 + \Gamma_1)(w_2+\gamma_2 + \Gamma_2)+w_1(w_2+\gamma_2 + \Gamma_2) + w_2(w_1+\gamma_1 + \Gamma_1)}{w_1\gamma_1(w_2+\gamma_2 + \Gamma_2)+w_2\gamma_2(w_1+\gamma_1 + \Gamma_1)} 
\ee 
The dependence of $\tau$ on the detuning from level crossing $\delta\epsilon=\epsilon_1-\epsilon_2$ can be analyzed using the expressions for the rotation matrix (Eq.~2 in the main text), giving 
\bea 
&& \gamma_1/\gamma=\sin^2\theta 
,\quad 
\gamma_2/\gamma=\cos^2\theta 
, 
\\ 
&& w_1/W=\cos^2\theta 
,\quad 
w_2/W=\sin^2\theta 
, 
\eea 
with  
\be 
\label{cos2theta}\cos 2\theta=\delta\epsilon/\sqrt{\delta\epsilon^2+\Delta^2}. 
\ee 
 
Substituting these expression into the equation for $\tau$, we arrive at 
\be\label{tau1} 
\tau= 
\frac{\left(3\frac{W}{\gamma} + \frac{\gamma}{W} + 4 + 8\frac{\delta\epsilon^2}{\Delta^2}\right) + \frac{4}{\sin^2 2\theta}\left(2\frac{\Gamma_1}{\gamma}\sin^2\theta + 2\frac{\Gamma_2}{\gamma}\cos^2\theta + \frac{\Gamma_1}{W}\cos^2\theta + \frac{\Gamma_2}{W}\sin^2\theta + \frac{\Gamma_1\Gamma_2}{\gamma W}\right)} 
{W + \gamma + \Gamma_1 + \Gamma_2} 
, 
\ee 
where $\cos^2 \theta = \frac12(1 + \cos 2\theta)$ and $\sin^2\theta = \frac12(1 - \cos 2\theta)$, with $\cos 2\theta$ defined in Eq.(\ref{cos2theta}). The result (\ref{tau1}) generalizes the simplified model discussed in the main text, in which $\Gamma_1$ and $\Gamma_2$ were assumed to be small compared to $\gamma$ and $W$, and therefore ignored.

Expression (\ref{tau1}), where $\Gamma_{1,2}$ are allowed to have arbitrary values and in principle arbitrary energy dependence, is rather complicated. 
For simplicity, we now take $\Gamma_1 = \Gamma_2$, independent of energy, and find a Lorentzian dependence: 
\be\label{eq:GammaSM} 
\Gamma(\delta\epsilon)=\frac{\Delta^2 (W+\gamma + 2\Gamma_1)}{\left[8 + 4\left(2\frac{\Gamma_1}{\gamma} + \frac{\Gamma_1}{W} + \frac{\Gamma_1^2}{\gamma W}\right)\right](\delta\epsilon^2+\gamma_*^2)},\quad 
\gamma_*^2 = \frac{\Delta^2\left(3\frac{W}{\gamma} + \frac{\gamma}{W} + 4\right) + 8\frac{\Gamma_1}{\gamma} + 4\frac{\Gamma_1}{W} + 4\frac{\Gamma_1^2}{\gamma W}  }{8+ 4\left(2\frac{\Gamma_1}{\gamma} + \frac{\Gamma_1}{W} + \frac{\Gamma_1^2}{\gamma W}\right)}. 
\ee 
The width $\gamma_*$ of the Lorentzian (\ref{eq:GammaSM}) now has a more complicated dependence on parameters than in the absence of relaxation.   
Solving for the minimum width, found by setting $d\gamma_*/dW = 0$, requires finding the roots of a cubic polynomial. The analysis shows that the dependence of the width $\gamma_*$ on the excitation strength $W$ is nonmonotonic, reproducing the behavior discussed in the main text, with the narrowest resonance width attained at some finite value of $W$. 
We note that $\Gamma_1$ provides a cutoff at small $W$,  
so that the width of the resonance no longer diverges for small $W$. 

It is important to note that a simple Fermi's Golden Rule (FGR) calculation of the direct ionization rate of the ground state, which does not account for population build-up in the excited states, fails to explain the observed behavior. After integrating the FGR ionization rate over the broadband spectrum of $V(t)$, we obtain a transition rate which is {\it independent} of the detuning from resonance.  
Thus taking into account the bottleneck effect in the rate equations is essential for understanding the enhancement of the ionization rate near resonance. 

\end{widetext} 

\bibliographystyle{apsrev4-1} 
\bibliography{RingSpectroscopy}

\begin{thebibliography}{29}%
\makeatletter
\providecommand \@ifxundefined [1]{%
 \@ifx{#1\undefined}
}%
\providecommand \@ifnum [1]{%
 \ifnum #1\expandafter \@firstoftwo
 \else \expandafter \@secondoftwo
 \fi
}%
\providecommand \@ifx [1]{%
 \ifx #1\expandafter \@firstoftwo
 \else \expandafter \@secondoftwo
 \fi
}%
\providecommand \natexlab [1]{#1}%
\providecommand \enquote  [1]{``#1''}%
\providecommand \bibnamefont  [1]{#1}%
\providecommand \bibfnamefont [1]{#1}%
\providecommand \citenamefont [1]{#1}%
\providecommand \href@noop [0]{\@secondoftwo}%
\providecommand \href [0]{\begingroup \@sanitize@url \@href}%
\providecommand \@href[1]{\@@startlink{#1}\@@href}%
\providecommand \@@href[1]{\endgroup#1\@@endlink}%
\providecommand \@sanitize@url [0]{\catcode `\\12\catcode `\$12\catcode
  `\&12\catcode `\#12\catcode `\^12\catcode `\_12\catcode `\%12\relax}%
\providecommand \@@startlink[1]{}%
\providecommand \@@endlink[0]{}%
\providecommand \url  [0]{\begingroup\@sanitize@url \@url }%
\providecommand \@url [1]{\endgroup\@href {#1}{\urlprefix }}%
\providecommand \urlprefix  [0]{URL }%
\providecommand \Eprint [0]{\href }%
\providecommand \doibase [0]{http://dx.doi.org/}%
\providecommand \selectlanguage [0]{\@gobble}%
\providecommand \bibinfo  [0]{\@secondoftwo}%
\providecommand \bibfield  [0]{\@secondoftwo}%
\providecommand \translation [1]{[#1]}%
\providecommand \BibitemOpen [0]{}%
\providecommand \bibitemStop [0]{}%
\providecommand \bibitemNoStop [0]{.\EOS\space}%
\providecommand \EOS [0]{\spacefactor3000\relax}%
\providecommand \BibitemShut  [1]{\csname bibitem#1\endcsname}%
\let\auto@bib@innerbib\@empty
\bibitem [{\citenamefont {Field}\ \emph {et~al.}(1993)\citenamefont {Field},
  \citenamefont {Smith}, \citenamefont {Pepper}, \citenamefont {Ritchie},
  \citenamefont {Frost}, \citenamefont {Jones},\ and\ \citenamefont
  {Hasko}}]{field:1993}%
  \BibitemOpen
  \bibfield  {author} {\bibinfo {author} {\bibfnamefont {M.}~\bibnamefont
  {Field}}, \bibinfo {author} {\bibfnamefont {C.~G.}\ \bibnamefont {Smith}},
  \bibinfo {author} {\bibfnamefont {M.}~\bibnamefont {Pepper}}, \bibinfo
  {author} {\bibfnamefont {D.~A.}\ \bibnamefont {Ritchie}}, \bibinfo {author}
  {\bibfnamefont {J.~E.~F.}\ \bibnamefont {Frost}}, \bibinfo {author}
  {\bibfnamefont {G.~A.~C.}\ \bibnamefont {Jones}}, \ and\ \bibinfo {author}
  {\bibfnamefont {D.~G.}\ \bibnamefont {Hasko}},\ }\href@noop {} {\bibfield
  {journal} {\bibinfo  {journal} {Phys. Rev. Lett.}\ }\textbf {\bibinfo
  {volume} {70}},\ \bibinfo {pages} {1311} (\bibinfo {year}
  {1993})}\BibitemShut {NoStop}%
\bibitem [{\citenamefont {Aguado}\ and\ \citenamefont
  {Kouwenhoven}(2000)}]{aguado:2000}%
  \BibitemOpen
  \bibfield  {author} {\bibinfo {author} {\bibfnamefont {R.}~\bibnamefont
  {Aguado}}\ and\ \bibinfo {author} {\bibfnamefont {L.~P.}\ \bibnamefont
  {Kouwenhoven}},\ }\href@noop {} {\bibfield  {journal} {\bibinfo  {journal}
  {Phys. Rev. Lett.}\ }\textbf {\bibinfo {volume} {84}},\ \bibinfo {pages}
  {001986} (\bibinfo {year} {2000})}\BibitemShut {NoStop}%
\bibitem [{\citenamefont {Onac}\ \emph {et~al.}(2006)\citenamefont {Onac},
  \citenamefont {Balestro}, \citenamefont {{van Beveren}}, \citenamefont
  {Hartmann}, \citenamefont {Nazarov},\ and\ \citenamefont
  {Kouwenhoven}}]{onacQD:2006}%
  \BibitemOpen
  \bibfield  {author} {\bibinfo {author} {\bibfnamefont {E.}~\bibnamefont
  {Onac}}, \bibinfo {author} {\bibfnamefont {F.}~\bibnamefont {Balestro}},
  \bibinfo {author} {\bibfnamefont {L.~H.~W.}\ \bibnamefont {{van Beveren}}},
  \bibinfo {author} {\bibfnamefont {U.}~\bibnamefont {Hartmann}}, \bibinfo
  {author} {\bibfnamefont {Y.~V.}\ \bibnamefont {Nazarov}}, \ and\ \bibinfo
  {author} {\bibfnamefont {L.~P.}\ \bibnamefont {Kouwenhoven}},\ }\href@noop {}
  {\bibfield  {journal} {\bibinfo  {journal} {Phys. Rev. Lett.}\ }\textbf
  {\bibinfo {volume} {96}},\ \bibinfo {pages} {176601} (\bibinfo {year}
  {2006})}\BibitemShut {NoStop}%
\bibitem [{\citenamefont {Gustavsson}\ \emph {et~al.}(2007)\citenamefont
  {Gustavsson}, \citenamefont {Studer}, \citenamefont {Leturcq}, \citenamefont
  {Ihn}, \citenamefont {Ensslin}, \citenamefont {Driscoll},\ and\ \citenamefont
  {Gossard}}]{gustavssonPRL:2007}%
  \BibitemOpen
  \bibfield  {author} {\bibinfo {author} {\bibfnamefont {S.}~\bibnamefont
  {Gustavsson}}, \bibinfo {author} {\bibfnamefont {M.}~\bibnamefont {Studer}},
  \bibinfo {author} {\bibfnamefont {R.}~\bibnamefont {Leturcq}}, \bibinfo
  {author} {\bibfnamefont {T.}~\bibnamefont {Ihn}}, \bibinfo {author}
  {\bibfnamefont {K.}~\bibnamefont {Ensslin}}, \bibinfo {author} {\bibfnamefont
  {D.~C.}\ \bibnamefont {Driscoll}}, \ and\ \bibinfo {author} {\bibfnamefont
  {A.~C.}\ \bibnamefont {Gossard}},\ }\href@noop {} {\bibfield  {journal}
  {\bibinfo  {journal} {Phys. Rev. Lett.}\ }\textbf {\bibinfo {volume} {99}},\
  \bibinfo {pages} {206804} (\bibinfo {year} {2007})}\BibitemShut {NoStop}%
\bibitem [{\citenamefont {Vandersypen}\ \emph {et~al.}(2004)\citenamefont
  {Vandersypen}, \citenamefont {Elzerman}, \citenamefont {Schouten},
  \citenamefont {Willems~van Beveren}, \citenamefont {Hanson},\ and\
  \citenamefont {Kouwenhoven}}]{vandersypen:2004}%
  \BibitemOpen
  \bibfield  {author} {\bibinfo {author} {\bibfnamefont {L.~M.~K.}\
  \bibnamefont {Vandersypen}}, \bibinfo {author} {\bibfnamefont {J.~M.}\
  \bibnamefont {Elzerman}}, \bibinfo {author} {\bibfnamefont {R.~N.}\
  \bibnamefont {Schouten}}, \bibinfo {author} {\bibfnamefont {L.~H.}\
  \bibnamefont {Willems~van Beveren}}, \bibinfo {author} {\bibfnamefont
  {R.}~\bibnamefont {Hanson}}, \ and\ \bibinfo {author} {\bibfnamefont {L.~P.}\
  \bibnamefont {Kouwenhoven}},\ }\href@noop {} {\bibfield  {journal} {\bibinfo
  {journal} {Appl. Phys. Lett.}\ }\textbf {\bibinfo {volume} {85}},\ \bibinfo
  {pages} {4394} (\bibinfo {year} {2004})}\BibitemShut {NoStop}%
\bibitem [{\citenamefont {Schleser}\ \emph {et~al.}(2004)\citenamefont
  {Schleser}, \citenamefont {Ruh}, \citenamefont {Ihn}, \citenamefont
  {Ensslin}, \citenamefont {Driscoll},\ and\ \citenamefont
  {Gossard}}]{schleser:2004}%
  \BibitemOpen
  \bibfield  {author} {\bibinfo {author} {\bibfnamefont {R.}~\bibnamefont
  {Schleser}}, \bibinfo {author} {\bibfnamefont {E.}~\bibnamefont {Ruh}},
  \bibinfo {author} {\bibfnamefont {T.}~\bibnamefont {Ihn}}, \bibinfo {author}
  {\bibfnamefont {K.}~\bibnamefont {Ensslin}}, \bibinfo {author} {\bibfnamefont
  {D.~C.}\ \bibnamefont {Driscoll}}, \ and\ \bibinfo {author} {\bibfnamefont
  {A.~C.}\ \bibnamefont {Gossard}},\ }\href@noop {} {\bibfield  {journal}
  {\bibinfo  {journal} {Appl. Phys. Lett.}\ }\textbf {\bibinfo {volume} {85}},\
  \bibinfo {pages} {2005} (\bibinfo {year} {2004})}\BibitemShut {NoStop}%
\bibitem [{\citenamefont {Fujisawa}\ \emph {et~al.}(2004)\citenamefont
  {Fujisawa}, \citenamefont {Hayashi}, \citenamefont {Hirayama}, \citenamefont
  {Cheong},\ and\ \citenamefont {Jeong}}]{fujisawa:2004}%
  \BibitemOpen
  \bibfield  {author} {\bibinfo {author} {\bibfnamefont {T.}~\bibnamefont
  {Fujisawa}}, \bibinfo {author} {\bibfnamefont {T.}~\bibnamefont {Hayashi}},
  \bibinfo {author} {\bibfnamefont {Y.}~\bibnamefont {Hirayama}}, \bibinfo
  {author} {\bibfnamefont {H.~D.}\ \bibnamefont {Cheong}}, \ and\ \bibinfo
  {author} {\bibfnamefont {Y.~H.}\ \bibnamefont {Jeong}},\ }\href@noop {}
  {\bibfield  {journal} {\bibinfo  {journal} {Appl. Phys. Lett.}\ }\textbf
  {\bibinfo {volume} {84}},\ \bibinfo {pages} {2343} (\bibinfo {year}
  {2004})}\BibitemShut {NoStop}%
\bibitem [{\citenamefont {Elzerman}\ \emph {et~al.}(2004)\citenamefont
  {Elzerman}, \citenamefont {Hanson}, \citenamefont {Willems~van Beveren},
  \citenamefont {Witkamp}, \citenamefont {Vandersypen},\ and\ \citenamefont
  {Kouwenhoven}}]{elzermanNature:2004}%
  \BibitemOpen
  \bibfield  {author} {\bibinfo {author} {\bibfnamefont {J.~M.}\ \bibnamefont
  {Elzerman}}, \bibinfo {author} {\bibfnamefont {R.}~\bibnamefont {Hanson}},
  \bibinfo {author} {\bibfnamefont {L.~H.}\ \bibnamefont {Willems~van
  Beveren}}, \bibinfo {author} {\bibfnamefont {B.}~\bibnamefont {Witkamp}},
  \bibinfo {author} {\bibfnamefont {L.~M.~K.}\ \bibnamefont {Vandersypen}}, \
  and\ \bibinfo {author} {\bibfnamefont {L.~P.}\ \bibnamefont {Kouwenhoven}},\
  }\href@noop {} {\bibfield  {journal} {\bibinfo  {journal} {Nature}\ }\textbf
  {\bibinfo {volume} {430}},\ \bibinfo {pages} {431} (\bibinfo {year}
  {2004})}\BibitemShut {NoStop}%
\bibitem [{\citenamefont {Amasha}\ \emph {et~al.}(2008)\citenamefont {Amasha},
  \citenamefont {MacLean}, \citenamefont {Radu}, \citenamefont {Zumbuhl},
  \citenamefont {Kastner}, \citenamefont {Hanson},\ and\ \citenamefont
  {Gossard}}]{amasha:2008}%
  \BibitemOpen
  \bibfield  {author} {\bibinfo {author} {\bibfnamefont {S.}~\bibnamefont
  {Amasha}}, \bibinfo {author} {\bibfnamefont {K.}~\bibnamefont {MacLean}},
  \bibinfo {author} {\bibfnamefont {I.~P.}\ \bibnamefont {Radu}}, \bibinfo
  {author} {\bibfnamefont {D.~M.}\ \bibnamefont {Zumbuhl}}, \bibinfo {author}
  {\bibfnamefont {M.~A.}\ \bibnamefont {Kastner}}, \bibinfo {author}
  {\bibfnamefont {M.~P.}\ \bibnamefont {Hanson}}, \ and\ \bibinfo {author}
  {\bibfnamefont {A.~C.}\ \bibnamefont {Gossard}},\ }\href@noop {} {\bibfield
  {journal} {\bibinfo  {journal} {Phys. Rev. Lett.}\ }\textbf {\bibinfo
  {volume} {100}},\ \bibinfo {pages} {046803} (\bibinfo {year}
  {2008})}\BibitemShut {NoStop}%
\bibitem [{\citenamefont {Flindt}\ \emph {et~al.}(2009)\citenamefont {Flindt},
  \citenamefont {Fricke}, \citenamefont {Hohls}, \citenamefont {Novotny},
  \citenamefont {Netocny}, \citenamefont {Brandes},\ and\ \citenamefont
  {Haug}}]{Flindt:2009}%
  \BibitemOpen
  \bibfield  {author} {\bibinfo {author} {\bibfnamefont {C.}~\bibnamefont
  {Flindt}}, \bibinfo {author} {\bibfnamefont {C.}~\bibnamefont {Fricke}},
  \bibinfo {author} {\bibfnamefont {F.}~\bibnamefont {Hohls}}, \bibinfo
  {author} {\bibfnamefont {T.}~\bibnamefont {Novotny}}, \bibinfo {author}
  {\bibfnamefont {K.}~\bibnamefont {Netocny}}, \bibinfo {author} {\bibfnamefont
  {T.}~\bibnamefont {Brandes}}, \ and\ \bibinfo {author} {\bibfnamefont
  {R.~J.}\ \bibnamefont {Haug}},\ }\href {\doibase 10.1073/pnas.0901002106}
  {\bibfield  {journal} {\bibinfo  {journal} {PNAS}\ }\textbf {\bibinfo
  {volume} {106}},\ \bibinfo {pages} {10116} (\bibinfo {year}
  {2009})}\BibitemShut {NoStop}%
\bibitem [{\citenamefont {Madden}\ and\ \citenamefont
  {Codling}(1963)}]{madden:1963}%
  \BibitemOpen
  \bibfield  {author} {\bibinfo {author} {\bibfnamefont {R.~P.}\ \bibnamefont
  {Madden}}\ and\ \bibinfo {author} {\bibfnamefont {K.}~\bibnamefont
  {Codling}},\ }\href@noop {} {\bibfield  {journal} {\bibinfo  {journal} {Phys.
  Rev. Lett.}\ }\textbf {\bibinfo {volume} {10}},\ \bibinfo {pages} {516}
  (\bibinfo {year} {1963})}\BibitemShut {NoStop}%
\bibitem [{\citenamefont {Biondi}\ \emph {et~al.}(1979)\citenamefont {Biondi},
  \citenamefont {Herzenberg},\ and\ \citenamefont {Kuyatt}}]{biondi:1979}%
  \BibitemOpen
  \bibfield  {author} {\bibinfo {author} {\bibfnamefont {M.~A.}\ \bibnamefont
  {Biondi}}, \bibinfo {author} {\bibfnamefont {A.}~\bibnamefont {Herzenberg}},
  \ and\ \bibinfo {author} {\bibfnamefont {C.~E.}\ \bibnamefont {Kuyatt}},\
  }\href@noop {} {\bibfield  {journal} {\bibinfo  {journal} {Physics Today}\
  }\textbf {\bibinfo {volume} {32}},\ \bibinfo {pages} {44} (\bibinfo {year}
  {1979})}\BibitemShut {NoStop}%
\bibitem [{\citenamefont {Madden}\ and\ \citenamefont
  {Parr}(1982)}]{madden:1982}%
  \BibitemOpen
  \bibfield  {author} {\bibinfo {author} {\bibfnamefont {R.~P.}\ \bibnamefont
  {Madden}}\ and\ \bibinfo {author} {\bibfnamefont {A.~G.}\ \bibnamefont
  {Parr}},\ }\href@noop {} {\bibfield  {journal} {\bibinfo  {journal} {Appl.
  Opt.}\ }\textbf {\bibinfo {volume} {21}},\ \bibinfo {pages} {179} (\bibinfo
  {year} {1982})}\BibitemShut {NoStop}%
\bibitem [{\citenamefont {Blick}\ \emph {et~al.}(1995)\citenamefont {Blick},
  \citenamefont {Haug}, \citenamefont {van~der Weide}, \citenamefont {von
  Klitzing},\ and\ \citenamefont {Eberl}}]{blick:1995}%
  \BibitemOpen
  \bibfield  {author} {\bibinfo {author} {\bibfnamefont {R.~H.}\ \bibnamefont
  {Blick}}, \bibinfo {author} {\bibfnamefont {R.~J.}\ \bibnamefont {Haug}},
  \bibinfo {author} {\bibfnamefont {D.~W.}\ \bibnamefont {van~der Weide}},
  \bibinfo {author} {\bibfnamefont {K.}~\bibnamefont {von Klitzing}}, \ and\
  \bibinfo {author} {\bibfnamefont {K.}~\bibnamefont {Eberl}},\ }\href@noop {}
  {\bibfield  {journal} {\bibinfo  {journal} {Appl. Phys. Lett.}\ }\textbf
  {\bibinfo {volume} {67}},\ \bibinfo {pages} {3924} (\bibinfo {year}
  {1995})}\BibitemShut {NoStop}%
\bibitem [{\citenamefont {Oosterkamp}\ \emph {et~al.}(1997)\citenamefont
  {Oosterkamp}, \citenamefont {Kouwenhoven}, \citenamefont {Koolen},
  \citenamefont {van~der Vaart},\ and\ \citenamefont
  {Harmans}}]{oosterkamp:1997}%
  \BibitemOpen
  \bibfield  {author} {\bibinfo {author} {\bibfnamefont {T.~H.}\ \bibnamefont
  {Oosterkamp}}, \bibinfo {author} {\bibfnamefont {L.~P.}\ \bibnamefont
  {Kouwenhoven}}, \bibinfo {author} {\bibfnamefont {A.~E.~A.}\ \bibnamefont
  {Koolen}}, \bibinfo {author} {\bibfnamefont {N.~C.}\ \bibnamefont {van~der
  Vaart}}, \ and\ \bibinfo {author} {\bibfnamefont {C.~J. P.~M.}\ \bibnamefont
  {Harmans}},\ }\href@noop {} {\bibfield  {journal} {\bibinfo  {journal} {Phys.
  Rev. Lett.}\ }\textbf {\bibinfo {volume} {78}},\ \bibinfo {pages} {1536}
  (\bibinfo {year} {1997})}\BibitemShut {NoStop}%
\bibitem [{\citenamefont {Holleitner}\ \emph {et~al.}(2002)\citenamefont
  {Holleitner}, \citenamefont {Blick}, \citenamefont {Huttel}, \citenamefont
  {Eberl},\ and\ \citenamefont {Kotthaus}}]{Holleitner:2002}%
  \BibitemOpen
  \bibfield  {author} {\bibinfo {author} {\bibfnamefont {A.~W.}\ \bibnamefont
  {Holleitner}}, \bibinfo {author} {\bibfnamefont {R.~H.}\ \bibnamefont
  {Blick}}, \bibinfo {author} {\bibfnamefont {A.~K.}\ \bibnamefont {Huttel}},
  \bibinfo {author} {\bibfnamefont {K.}~\bibnamefont {Eberl}}, \ and\ \bibinfo
  {author} {\bibfnamefont {J.~P.}\ \bibnamefont {Kotthaus}},\ }\href {\doibase
  10.1126/science.1071215} {\bibfield  {journal} {\bibinfo  {journal}
  {Science}\ }\textbf {\bibinfo {volume} {297}},\ \bibinfo {pages} {70}
  (\bibinfo {year} {2002})}\BibitemShut {NoStop}%
\bibitem [{\citenamefont {van~der Wiel}\ \emph {et~al.}(2003)\citenamefont
  {van~der Wiel}, \citenamefont {De~Franceschi}, \citenamefont {Elzerman},
  \citenamefont {Fujisawa}, \citenamefont {Tarucha},\ and\ \citenamefont
  {Kouwenhoven}}]{vanderwiel:2003}%
  \BibitemOpen
  \bibfield  {author} {\bibinfo {author} {\bibfnamefont {W.~G.}\ \bibnamefont
  {van~der Wiel}}, \bibinfo {author} {\bibfnamefont {S.}~\bibnamefont
  {De~Franceschi}}, \bibinfo {author} {\bibfnamefont {J.~M.}\ \bibnamefont
  {Elzerman}}, \bibinfo {author} {\bibfnamefont {T.}~\bibnamefont {Fujisawa}},
  \bibinfo {author} {\bibfnamefont {S.}~\bibnamefont {Tarucha}}, \ and\
  \bibinfo {author} {\bibfnamefont {L.~P.}\ \bibnamefont {Kouwenhoven}},\
  }\href {\doibase 10.1103/RevModPhys.75.1} {\bibfield  {journal} {\bibinfo
  {journal} {Rev. Mod. Phys.}\ }\textbf {\bibinfo {volume} {75}},\ \bibinfo
  {pages} {1} (\bibinfo {year} {2003})}\BibitemShut {NoStop}%
\bibitem [{\citenamefont {Devoret}\ \emph {et~al.}(1990)\citenamefont
  {Devoret}, \citenamefont {Esteve}, \citenamefont {Grabert}, \citenamefont
  {Ingold}, \citenamefont {Pothier},\ and\ \citenamefont
  {Urbina}}]{devoret:1990}%
  \BibitemOpen
  \bibfield  {author} {\bibinfo {author} {\bibfnamefont {M.~H.}\ \bibnamefont
  {Devoret}}, \bibinfo {author} {\bibfnamefont {D.}~\bibnamefont {Esteve}},
  \bibinfo {author} {\bibfnamefont {H.}~\bibnamefont {Grabert}}, \bibinfo
  {author} {\bibfnamefont {G.~L.}\ \bibnamefont {Ingold}}, \bibinfo {author}
  {\bibfnamefont {H.}~\bibnamefont {Pothier}}, \ and\ \bibinfo {author}
  {\bibfnamefont {C.}~\bibnamefont {Urbina}},\ }\href@noop {} {\bibfield
  {journal} {\bibinfo  {journal} {Phys. Rev. Lett.}\ }\textbf {\bibinfo
  {volume} {64}},\ \bibinfo {pages} {1824} (\bibinfo {year}
  {1990})}\BibitemShut {NoStop}%
\bibitem [{\citenamefont {Girvin}\ \emph {et~al.}(1990)\citenamefont {Girvin},
  \citenamefont {Glazman}, \citenamefont {Jonson}, \citenamefont {Penn},\ and\
  \citenamefont {Stiles}}]{girvin:1990}%
  \BibitemOpen
  \bibfield  {author} {\bibinfo {author} {\bibfnamefont {S.~M.}\ \bibnamefont
  {Girvin}}, \bibinfo {author} {\bibfnamefont {L.~I.}\ \bibnamefont {Glazman}},
  \bibinfo {author} {\bibfnamefont {M.}~\bibnamefont {Jonson}}, \bibinfo
  {author} {\bibfnamefont {D.~R.}\ \bibnamefont {Penn}}, \ and\ \bibinfo
  {author} {\bibfnamefont {M.~D.}\ \bibnamefont {Stiles}},\ }\href@noop {}
  {\bibfield  {journal} {\bibinfo  {journal} {Phys. Rev. Lett.}\ }\textbf
  {\bibinfo {volume} {64}},\ \bibinfo {pages} {3183} (\bibinfo {year}
  {1990})}\BibitemShut {NoStop}%
\bibitem [{\citenamefont {Gustavsson}\ \emph
  {et~al.}(2008{\natexlab{a}})\citenamefont {Gustavsson}, \citenamefont
  {Shorubalko}, \citenamefont {Leturcq}, \citenamefont {Ihn}, \citenamefont
  {Ensslin},\ and\ \citenamefont {Schoen}}]{gustavssonNWPRB:2008}%
  \BibitemOpen
  \bibfield  {author} {\bibinfo {author} {\bibfnamefont {S.}~\bibnamefont
  {Gustavsson}}, \bibinfo {author} {\bibfnamefont {I.}~\bibnamefont
  {Shorubalko}}, \bibinfo {author} {\bibfnamefont {R.}~\bibnamefont {Leturcq}},
  \bibinfo {author} {\bibfnamefont {T.}~\bibnamefont {Ihn}}, \bibinfo {author}
  {\bibfnamefont {K.}~\bibnamefont {Ensslin}}, \ and\ \bibinfo {author}
  {\bibfnamefont {S.}~\bibnamefont {Schoen}},\ }\href {\doibase
  10.1103/PhysRevB.78.035324} {\bibfield  {journal} {\bibinfo  {journal} {Phys.
  Rev. B}\ }\textbf {\bibinfo {volume} {78}},\ \bibinfo {pages} {035324}
  (\bibinfo {year} {2008}{\natexlab{a}})}\BibitemShut {NoStop}%
\bibitem [{\citenamefont {Kouwenhoven}\ \emph {et~al.}(1997)\citenamefont
  {Kouwenhoven}, \citenamefont {Marcus}, \citenamefont {McEuen}, \citenamefont
  {Tarucha}, \citenamefont {Westervelt},\ and\ \citenamefont
  {Wingreen}}]{kouwenhoven:1997}%
  \BibitemOpen
  \bibfield  {author} {\bibinfo {author} {\bibfnamefont {L.~P.}\ \bibnamefont
  {Kouwenhoven}}, \bibinfo {author} {\bibfnamefont {C.~M.}\ \bibnamefont
  {Marcus}}, \bibinfo {author} {\bibfnamefont {P.~M.}\ \bibnamefont {McEuen}},
  \bibinfo {author} {\bibfnamefont {S.}~\bibnamefont {Tarucha}}, \bibinfo
  {author} {\bibfnamefont {R.~M.}\ \bibnamefont {Westervelt}}, \ and\ \bibinfo
  {author} {\bibfnamefont {N.~S.}\ \bibnamefont {Wingreen}},\ }in\ \href@noop
  {} {\emph {\bibinfo {booktitle} {Mesoscopic Electron Transport}}},\ \bibinfo
  {series and number} {NATO ASI Ser. E 345},\ \bibinfo {editor} {edited by\
  \bibinfo {editor} {\bibfnamefont {L.~L.}\ \bibnamefont {Sohn}}, \bibinfo
  {editor} {\bibfnamefont {L.~P.}\ \bibnamefont {Kouwenhoven}}, \ and\ \bibinfo
  {editor} {\bibfnamefont {G.}~\bibnamefont {Sch\"on}}}\ (\bibinfo  {publisher}
  {Kluwer},\ \bibinfo {address} {Dordrecht},\ \bibinfo {year} {1997})\ pp.\
  \bibinfo {pages} {105--214}\BibitemShut {NoStop}%
\bibitem [{\citenamefont {Cronenwett}\ \emph {et~al.}(1997)\citenamefont
  {Cronenwett}, \citenamefont {Patel}, \citenamefont {Marcus}, \citenamefont
  {Campman},\ and\ \citenamefont {Gossard}}]{cronenwett:1997}%
  \BibitemOpen
  \bibfield  {author} {\bibinfo {author} {\bibfnamefont {S.~M.}\ \bibnamefont
  {Cronenwett}}, \bibinfo {author} {\bibfnamefont {S.~R.}\ \bibnamefont
  {Patel}}, \bibinfo {author} {\bibfnamefont {C.~M.}\ \bibnamefont {Marcus}},
  \bibinfo {author} {\bibfnamefont {K.}~\bibnamefont {Campman}}, \ and\
  \bibinfo {author} {\bibfnamefont {A.~C.}\ \bibnamefont {Gossard}},\
  }\href@noop {} {\bibfield  {journal} {\bibinfo  {journal} {Phys. Rev. Lett.}\
  }\textbf {\bibinfo {volume} {79}},\ \bibinfo {pages} {2312} (\bibinfo {year}
  {1997})}\BibitemShut {NoStop}%
\bibitem [{\citenamefont {Granger}\ \emph {et~al.}(2012)\citenamefont
  {Granger}, \citenamefont {Taubert}, \citenamefont {Young}, \citenamefont
  {Gaudreau}, \citenamefont {Kam}, \citenamefont {Studenikin}, \citenamefont
  {Zawadzki}, \citenamefont {Harbusch}, \citenamefont {Schuh}, \citenamefont
  {Wegscheider}, \citenamefont {Wasilewski}, \citenamefont {Clerk},
  \citenamefont {Ludwig},\ and\ \citenamefont {Sachrajda}}]{Granger:2012}%
  \BibitemOpen
  \bibfield  {author} {\bibinfo {author} {\bibfnamefont {G.}~\bibnamefont
  {Granger}}, \bibinfo {author} {\bibfnamefont {D.}~\bibnamefont {Taubert}},
  \bibinfo {author} {\bibfnamefont {C.~E.}\ \bibnamefont {Young}}, \bibinfo
  {author} {\bibfnamefont {L.}~\bibnamefont {Gaudreau}}, \bibinfo {author}
  {\bibfnamefont {A.}~\bibnamefont {Kam}}, \bibinfo {author} {\bibfnamefont
  {S.~A.}\ \bibnamefont {Studenikin}}, \bibinfo {author} {\bibfnamefont
  {P.}~\bibnamefont {Zawadzki}}, \bibinfo {author} {\bibfnamefont
  {D.}~\bibnamefont {Harbusch}}, \bibinfo {author} {\bibfnamefont
  {D.}~\bibnamefont {Schuh}}, \bibinfo {author} {\bibfnamefont
  {W.}~\bibnamefont {Wegscheider}}, \bibinfo {author} {\bibfnamefont {Z.~R.}\
  \bibnamefont {Wasilewski}}, \bibinfo {author} {\bibfnamefont {A.~A.}\
  \bibnamefont {Clerk}}, \bibinfo {author} {\bibfnamefont {S.}~\bibnamefont
  {Ludwig}}, \ and\ \bibinfo {author} {\bibfnamefont {A.~S.}\ \bibnamefont
  {Sachrajda}},\ }\href@noop {} {\bibfield  {journal} {\bibinfo  {journal}
  {Nature Physics}\ }\textbf {\bibinfo {volume} {8}},\ \bibinfo {pages} {522}
  (\bibinfo {year} {2012})}\BibitemShut {NoStop}%
\bibitem [{\citenamefont {Fujisawa}\ \emph {et~al.}(2002)\citenamefont
  {Fujisawa}, \citenamefont {Austing}, \citenamefont {Tokura}, \citenamefont
  {Hirayama},\ and\ \citenamefont {Tarucha}}]{fujisawa:2002}%
  \BibitemOpen
  \bibfield  {author} {\bibinfo {author} {\bibfnamefont {T.}~\bibnamefont
  {Fujisawa}}, \bibinfo {author} {\bibfnamefont {D.~G.}\ \bibnamefont
  {Austing}}, \bibinfo {author} {\bibfnamefont {Y.}~\bibnamefont {Tokura}},
  \bibinfo {author} {\bibfnamefont {Y.}~\bibnamefont {Hirayama}}, \ and\
  \bibinfo {author} {\bibfnamefont {S.}~\bibnamefont {Tarucha}},\ }\href@noop
  {} {\bibfield  {journal} {\bibinfo  {journal} {Nature}\ }\textbf {\bibinfo
  {volume} {419}},\ \bibinfo {pages} {278} (\bibinfo {year}
  {2002})}\BibitemShut {NoStop}%
\bibitem [{\citenamefont {Petta}\ \emph {et~al.}(2004)\citenamefont {Petta},
  \citenamefont {Johnson}, \citenamefont {Marcus}, \citenamefont {Hanson},\
  and\ \citenamefont {Gossard}}]{petta:2004}%
  \BibitemOpen
  \bibfield  {author} {\bibinfo {author} {\bibfnamefont {J.~R.}\ \bibnamefont
  {Petta}}, \bibinfo {author} {\bibfnamefont {A.~C.}\ \bibnamefont {Johnson}},
  \bibinfo {author} {\bibfnamefont {C.~M.}\ \bibnamefont {Marcus}}, \bibinfo
  {author} {\bibfnamefont {M.~P.}\ \bibnamefont {Hanson}}, \ and\ \bibinfo
  {author} {\bibfnamefont {A.~C.}\ \bibnamefont {Gossard}},\ }\href@noop {}
  {\bibfield  {journal} {\bibinfo  {journal} {Phys. Rev. Lett.}\ }\textbf
  {\bibinfo {volume} {93}},\ \bibinfo {pages} {186802} (\bibinfo {year}
  {2004})}\BibitemShut {NoStop}%
\bibitem [{\citenamefont {Fuhrer}\ \emph {et~al.}(2002)\citenamefont {Fuhrer},
  \citenamefont {Dorn}, \citenamefont {L\"uscher}, \citenamefont {Heinzel},
  \citenamefont {Ensslin}, \citenamefont {Wegscheider},\ and\ \citenamefont
  {Bichler}}]{Fuhrer:2004}%
  \BibitemOpen
  \bibfield  {author} {\bibinfo {author} {\bibfnamefont {A.}~\bibnamefont
  {Fuhrer}}, \bibinfo {author} {\bibfnamefont {A.}~\bibnamefont {Dorn}},
  \bibinfo {author} {\bibfnamefont {S.}~\bibnamefont {L\"uscher}}, \bibinfo
  {author} {\bibfnamefont {T.}~\bibnamefont {Heinzel}}, \bibinfo {author}
  {\bibfnamefont {K.}~\bibnamefont {Ensslin}}, \bibinfo {author} {\bibfnamefont
  {W.}~\bibnamefont {Wegscheider}}, \ and\ \bibinfo {author} {\bibfnamefont
  {M.}~\bibnamefont {Bichler}},\ }\href@noop {} {\bibfield  {journal} {\bibinfo
   {journal} {Superl. and Microstruc.}\ }\textbf {\bibinfo {volume} {31}},\
  \bibinfo {pages} {19} (\bibinfo {year} {2002})}\BibitemShut {NoStop}%
\bibitem [{\citenamefont {Gustavsson}\ \emph
  {et~al.}(2008{\natexlab{b}})\citenamefont {Gustavsson}, \citenamefont
  {Leturcq}, \citenamefont {Studer}, \citenamefont {Ihn}, \citenamefont
  {Ensslin}, \citenamefont {Driscoll},\ and\ \citenamefont
  {Gossard}}]{gustavssonNL:2008}%
  \BibitemOpen
  \bibfield  {author} {\bibinfo {author} {\bibfnamefont {S.}~\bibnamefont
  {Gustavsson}}, \bibinfo {author} {\bibfnamefont {R.}~\bibnamefont {Leturcq}},
  \bibinfo {author} {\bibfnamefont {M.}~\bibnamefont {Studer}}, \bibinfo
  {author} {\bibfnamefont {T.}~\bibnamefont {Ihn}}, \bibinfo {author}
  {\bibfnamefont {K.}~\bibnamefont {Ensslin}}, \bibinfo {author} {\bibfnamefont
  {D.~C.}\ \bibnamefont {Driscoll}}, \ and\ \bibinfo {author} {\bibfnamefont
  {A.~C.}\ \bibnamefont {Gossard}},\ }\href@noop {} {\bibfield  {journal}
  {\bibinfo  {journal} {Nano Letters}\ }\textbf {\bibinfo {volume} {8}},\
  \bibinfo {pages} {2547} (\bibinfo {year} {2008}{\natexlab{b}})}\BibitemShut
  {NoStop}%
\bibitem [{\citenamefont {Naaman}\ and\ \citenamefont
  {Aumentado}(2006)}]{naaman:2006}%
  \BibitemOpen
  \bibfield  {author} {\bibinfo {author} {\bibfnamefont {O.}~\bibnamefont
  {Naaman}}\ and\ \bibinfo {author} {\bibfnamefont {J.}~\bibnamefont
  {Aumentado}},\ }\href@noop {} {\bibfield  {journal} {\bibinfo  {journal}
  {Phys. Rev. Lett.}\ }\textbf {\bibinfo {volume} {96}},\ \bibinfo {pages}
  {100201} (\bibinfo {year} {2006})}\BibitemShut {NoStop}%
\bibitem [{\citenamefont {Fuhrer}\ \emph {et~al.}(2001)\citenamefont {Fuhrer},
  \citenamefont {Luescher}, \citenamefont {Ihn}, \citenamefont {Heinzel},
  \citenamefont {Ensslin}, \citenamefont {Wegscheider},\ and\ \citenamefont
  {Bichler}}]{fuhrer:2001}%
  \BibitemOpen
  \bibfield  {author} {\bibinfo {author} {\bibfnamefont {A.}~\bibnamefont
  {Fuhrer}}, \bibinfo {author} {\bibfnamefont {S.}~\bibnamefont {Luescher}},
  \bibinfo {author} {\bibfnamefont {T.}~\bibnamefont {Ihn}}, \bibinfo {author}
  {\bibfnamefont {T.}~\bibnamefont {Heinzel}}, \bibinfo {author} {\bibfnamefont
  {K.}~\bibnamefont {Ensslin}}, \bibinfo {author} {\bibfnamefont
  {W.}~\bibnamefont {Wegscheider}}, \ and\ \bibinfo {author} {\bibfnamefont
  {M.}~\bibnamefont {Bichler}},\ }\href@noop {} {\bibfield  {journal} {\bibinfo
   {journal} {Nature}\ }\textbf {\bibinfo {volume} {413}},\ \bibinfo {pages}
  {822} (\bibinfo {year} {2001})}\BibitemShut {NoStop}%
\end{thebibliography}%

\end{document}